\documentclass[sigconf]{acmart}

\AtBeginDocument{%
  }

\setcopyright{rightsretained}
\acmDOI{}
\acmConference[CHI '25 Workshop on Designing and Developing User Interfaces with AI]{}{April 27th, 2025}{Yokohama, Japan}
\acmISBN{}
\copyrightyear{2025}
\acmYear{2025}

\begin{document}
\emergencystretch 3em

\title{Expanding the Generative AI Design Space through Structured Prompting and Multimodal Interfaces}

\author{Nimisha Karnatak}
\authornote{Nimisha was working as a Student Researcher at Google Deepmind during this work.}

\affiliation{%
  \institution{University of Oxford}
  \city{Oxford}
  \country{UK}
}
\email{nimisha.karnatak@some.ox.ac.uk}

\author{Adrien Baranes}
\affiliation{%
  \institution{Google DeepMind}
    \country{London, UK}
}
\email{abaranes@google.com}

\author{Rob Marchant}
\affiliation{%
  \institution{Google DeepMind}
    \country{London, UK}
}
\email{rmarchant@google.com}

\author{Huinan Zeng}
\authornote{Huinan was working as a Student Researcher at Google Research during this work.}
\affiliation{%
  \institution{King's College London}
    \country{London, UK}
}
\email{huinan.zeng@kcl.ac.uk}

\author{Tríona Butler}
\affiliation{%
  \institution{Google DeepMind}
    \country{London, UK}
}
\email{trionab@google.com}

\author{Kristen Olson}
\affiliation{%
  \institution{Google DeepMind}
  \country{London, UK}
}
\email{kristenolson@google.com}

\renewcommand{\shortauthors}{Karnatak et al.}
\renewcommand{\shorttitle}{}

\settopmatter{printacmref=false}

\begin{abstract}
Text-based prompting remains the predominant interaction paradigm in generative AI, yet it often introduces friction for novice users such as small business owners (SBOs), who struggle to articulate creative goals in domain-specific contexts like advertising. Through a formative study with six SBOs in the United Kingdom, we identify three key challenges: difficulties in expressing brand intuition through prompts, limited opportunities for fine-grained adjustment and refinement during and after content generation, and the frequent production of generic content that lacks brand specificity. In response, we present ACAI (AI Co-Creation for Advertising and Inspiration), a multimodal generative AI tool designed to support novice designers by moving beyond traditional prompt interfaces. ACAI features a structured input system composed of three panels: Branding, Audience and Goals, and the Inspiration Board. These inputs allow users to convey brand-relevant context and visual preferences. This work contributes to HCI research on generative systems by showing how structured interfaces can foreground user-defined context, improve alignment, and enhance co-creative control in novice creative workflows.

\end{abstract}

\ccsdesc[500]{Human-centered computing~User interface design}
\ccsdesc[500]{Human-centered computing~Interaction design}
\ccsdesc[300]{Human-centered computing~Empirical studies in HCI}
\keywords{Structured Prompting, Interface Design, Novice Users, AI-Augmented Workflows, Multimodal Interfaces, GenAI}

\maketitle

\section{Introduction}

Small business owners (SBOs) represent the majority of business owners globally \cite{UN}. Visual advertising and branding are critical to the success of small enterprises; yet, many SBOs lack formal training in design and often do not have access to professional creative services~\cite{paradoxmarketingHatsWorn,bhalerao2022study}. The growing availability of generative AI systems presents new opportunities to support novice designers such as small business owners by reducing barriers to entry in advertising and branding and by broadening access to creative production tools.

To investigate how current generative systems support SBO creative workflows, we conducted a formative study with six small business owners (SBOs) in Manchester, United Kingdom. Using semi-structured interviews, we examined participants’ advertising practices, experiences with generative AI, and expectations for AI-assisted design. Our findings surfaced three recurring challenges: (1) the cognitive effort required to translate abstract, domain-specific goals into usable prompts; (2) dissatisfaction with generic outputs that failed to reflect brand tone, values, or audience needs; and (3) concerns about losing creative control due to over-automation, signaling the need for systems that enable co-creation and preserve user agency. These themes align with ongoing HCI work on promptability \cite{morris2024prompting} and the ``gulf of envisioning''~\cite{subramonyam2024bridging}, which highlight how non-expert users struggle to effectively guide generative systems.

To address this, we developed \textbf{ACAI} (AI Co-Creation for Advertising and Inspiration), a multimodal generative AI prototype that reimagines the prompt interface for novice designers. ACAI features a structured, panel-based user interface composed of three modules: the \textit{Branding \& Asset Panel}, the \textit{Audience \& Goals Panel}, and the \textit{Inspiration Board Panel}. These modules enable users to specify brand assets, define campaign objectives, and upload visual references. A multimodal large language model (MLLM) processes these structured inputs into a unified ``super prompt'' that guides advertisement generation aligned with user-defined goals. 

This work makes three primary contributions to HCI literature:
\vspace{-5pt}
\begin{itemize}
    \item[1.] A formative study surfacing promptability and brand-alignment challenges among SBOs using AI tools.
    \item[2.] Design requirements emphasizing multimodality and constraint-aware generation.
    \item[3.] The design of ACAI, a multimodal GenAI-powered prototype.
\end{itemize} 

\vspace{-9pt}

\section{Related Work}
We draw on two strands of prior research that shape our approach. The discussion below reviews foundational work in these areas and situates our contribution as an extension of ongoing efforts to design inclusive, context-aware generative systems.
\subsection{Contextualization in Generative and Interactive Systems}
Context-aware systems dynamically adapt to user needs by leveraging situational information captured through various modalities, such as GPS, sensors, and APIs~\cite{Dey2001Understanding,Chen2005ContextAware, Aliannejadi2021ContextAware, Clarizia2019Chatbot}. These systems have been widely adopted in domains such as mobile assistance and smart environments, where real-time responsiveness is essential. Building on this foundation, recent work has begun integrating context into generative workflows. For example, CARING-AI~\cite{shi2025caringai} generates spatially and temporally situated AR instructions using real-world environmental data, while ContextCam~\cite{fan2024contextcam} incorporates user-captured context to guide image generation. These systems demonstrate how contextual signals can increase semantic alignment and enhance user engagement.

However, current text-to-image generation models primarily rely on text-based prompts and offer limited affordances for encoding structured user intent or domain-specific constraints \cite{lee2023aligning}. While these systems cproduce compelling outputs, they often fall short in supporting intention-aligned co-creation, particularly for novice users in task-specific domains. Motivated by these limitations, we position ACAI to bridge the gap between generic text-to-image systems and context-sensitive creative workflows by foregrounding explicit, user-defined contextualization.

\subsection{Prompting Challenges and Interface Design for Novice Users}
A growing body of HCI research has highlighted the barriers novice users face when interacting with prompt-based generative systems. Subramonyam et al. ~\cite{subramonyam2024bridging} identify this challenge as the gulf of envisioning, which refers to a cognitive disconnect between what users intend to create and how they are able to articulate those intentions in forms that a system can interpret. Similarly, Zamfirescu-Pereira et al.~\cite{johnnycantprompt_nonAIexperts} demonstrate that non-AI experts frequently default to ineffective prompting patterns, revealing a lack of support for articulating goals in ways that models can reliably interpret. To address these challenges, recent work has explored interaction designs that simplify how users articulate intent. PromptCrafter~\cite{baek2023promptcrafter} introduces a dialogue-driven interface that helps users iteratively refine prompts with LLM support. These efforts reflect a broader shift in HCI toward designing promptable generative experiences, which are particularly effective in lowering entry barriers for novice and non-expert users.
Building on this trajectory, our work foregrounds small business owners (SBOs) as a critical yet underrepresented user group in HCI. By designing for their situated creative workflows and contextual constraints, we contribute to ongoing efforts to develop task-aligned generative systems that broaden the inclusivity and usability of promptable interfaces for non-expert users.

\begin{figure}[t]
    \centering
    \includegraphics[width=0.85\linewidth]{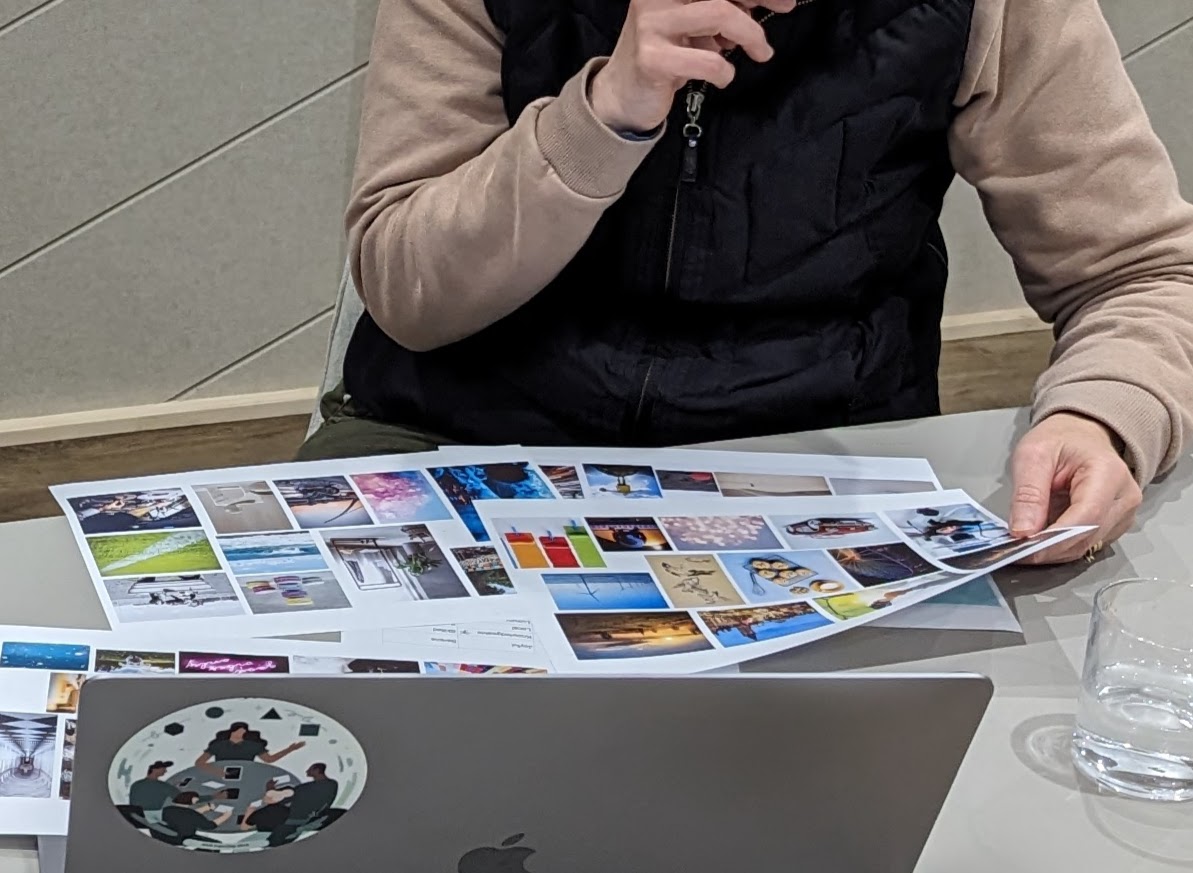} 
    \vspace{-6pt}
    \caption{Formative Study: 6 UK SBOs identify visual elements they would like to incorporate into their branding.\vspace{-10pt}}
    \label{fig:study}
\end{figure}

\section{Formative Study}
The last three authors conducted an in-person formative study with small business owners (SBOs) in Manchester, UK, to examine how SBOs perceive, engage with, and evaluate generative AI tools for advertising and brand development. The study aimed to understand existing branding practices, assess the alignment between generative AI value propositions and business needs, and identify adoption barriers. Participants were recruited through a professional research agency using an existing database of individuals open to participating in research. A screener questionnaire was distributed via email to identify eligible participants based on the following inclusion criteria: (1) over 18 years of age, (2) located in or near Manchester, and (3) a small business owner responsible for branding and the online presence of a business with a physical store or studio in the UK.

We conducted six in-person, semi-structured interviews, each lasting approximately 90 minutes. Sessions were audio recorded and transcribed for analysis. Participants received a £125 gift card as compensation. Businesses spanned diverse sectors, including food retail, entertainment, pet training, salon services, clothing retail, and personal services. Participants had varying levels of experience with generative AI tools, with most reporting some familiarity with platforms such as ChatGPT, Canva, or Gemini. Table~\ref{tab:participant_data} summarizes participant demographics and tool usage.

\subsection{Key Findings from Formative Research}

We analyzed the transcripts using reflexive thematic analysis. The following three findings emerged: 

\begin{itemize}

\item [1.] \textbf{Challenges in Articulating Brand Intuition through Prompts:} Participants struggled to articulate their brand vision using text-based prompts. While they possessed a strong intuitive understanding of their brand identity, they found it difficult to externalize this understanding. One participant shared, ``\textit{I instructed it to be funny, but that’s quite subjective, so I had to edit it quite heavily.” --P2}. This finding underscores the tacit and iterative nature of branding, highlighting how limitations in promptability introduce friction into the creative workflows of novice users.

\item [2.]\textbf{Generic and Misaligned AI Outputs:} Participants frequently described AI-generated advertisements as generic, repetitive, or misaligned with their brand’s tone and aesthetic. They noted that outputs often lacked the nuance necessary to reflect distinctive brand identities. For example, one participant reflected: \textit{“I was looking at some of these AI branding things, but nothing came of it, what I was offered... it was quite childish, quite gimmicky.” -- P1}. This disconnect reinforces the need for generative systems that incorporate user-defined brand elements and contextual cues to produce brand-consistent outputs.

\item [3.]\textbf{Tensions Between AI-Generated Content and User Control:} While participants appreciated the efficiency enabled by generative AI, many voiced concerns about single-prompt execution, where a text prompt yields a output that cannot be directly edited. Instead, users were required to craft new prompts to regenerate what they wanted, often through trial and error. Participants were not opposed to AI-generated content itself; rather, their frustration stemmed from outputs that misaligned with their intent and lacked accessible pathways for refinement. As one participant explained, ``\textit{You couldn’t really change it. When AI produced that for you, there weren’t tweaks you could make to it... so that’s the limitation of it” --P3}. These reflections underscore the importance of designing interaction models that support direct editability, iterative control, and co-creative engagement.

\end{itemize}

\begin{table}[h]
\centering
\begin{tabular}{|l|l|l|l|}
\hline
\textbf{Gender} & \textbf{Age} & \textbf{AI Tools} & \textbf{Business} \\ \hline
Man    & 31--40 & ChatGPT, Canva         & Food retail        \\ \hline
Man    & 31--40 & ChatGPT, Canva         & Entertainment      \\ \hline
Man    & 18--40 & ChatGPT, Gemini & Pet training       \\ \hline
Man    & 24--40 & Gemini\ & Salon             \\ \hline
Woman  & 41--50 & Canva & Clothing retail \\ \hline
Man    & 41--50 & ChatGPT, Canva            & Personal services  \\ \hline
\end{tabular}
\caption{Participant demographics and AI tool usage across business types}
\label{tab:participant_data}
\end{table}

\vspace{-10pt}

\subsection{Design Requirements}
Drawing on findings from our formative study, we outline design requirements to inform the development of generative AI systems that support small business owners in creating brand-aligned advertisements.

\begin{itemize}

  \item[1.] \textbf{DR1: Enable Multimodal Input and Interactive Style Selection for Creative Intent Expression}

  Participants demonstrated an intuitive understanding of their brand identity and a sense of which visual elements aligned with their advertisements. However, they lacked the formal design vocabulary to articulate these elements through text prompts alone. Many participants found free-form prompting unintuitive, often requiring iterative refinements or abandonment due to inability to easily edit. This mismatch led to frustration when AI-generated content did not meet expectations. To mitigate this:

  \begin{itemize}
    \item \textbf{Multimodal Input Affordances:} AI systems should allow users to specify creative intent through a combination of textual and visual modalities. Users should be able to upload or reference images alongside textual inputs, enabling a more comprehensive capture of their branding preferences.
    
    \item \textbf{Interactive Style Selection:} Tools should support users in extracting, refining, and integrating design attributes, such as fonts, color schemes, and composition, from reference visuals into AI-generated outputs. This functionality should work in tandem with text prompts to enhance expressivity and alignment with users' creative intent.
  \end{itemize}

  \item[2.] \textbf{DR2: Affordances and Constraints for Brand-Consistent AI Generation}

  Participants frequently expressed concerns that AI-generated advertisements did not adequately reflect their brand’s established identity, often resulting in generic and misaligned outputs. To ensure brand fidelity, generative systems should incorporate both expressive affordances and protective constraints:

  \begin{itemize}
    \item \textbf{Brand Identity Encoding Affordances:} AI tools should allow users to embed key elements of their brand identity, such as logos, specific fonts, color palettes, and tone of voice, into the generation pipeline. These affordances can mitigate users' concerns of generic outputs.

    \item \textbf{Brand Consistency Constraints:} The system should enforce constraints to preserve critical brand elements across outputs. These constraints act as fixed parameters (e.g., logo placement, business name inclusion, product-specific details) that ensure consistency while enabling flexibility across other design dimensions.

\end{itemize}

\end{itemize}

\section {ACAI: AI Co-Creation for Advertising and Inspiration}

In response to these requirements, we developed \textbf{ACAI (AI Co-Creation for Advertising and Inspiration)}, a generative AI-powered prototype designed to support the creation of brand-aligned advertisements. ACAI is designed for small business owners (SBOs) who possess deep contextual knowledge of their brand but lack formal design expertise. By aligning generative AI capabilities with users’ domain-specific insights, ACAI facilitates a co-creative design process.

\begin{figure}[h]
    \centering
    \includegraphics[width=0.85\linewidth]{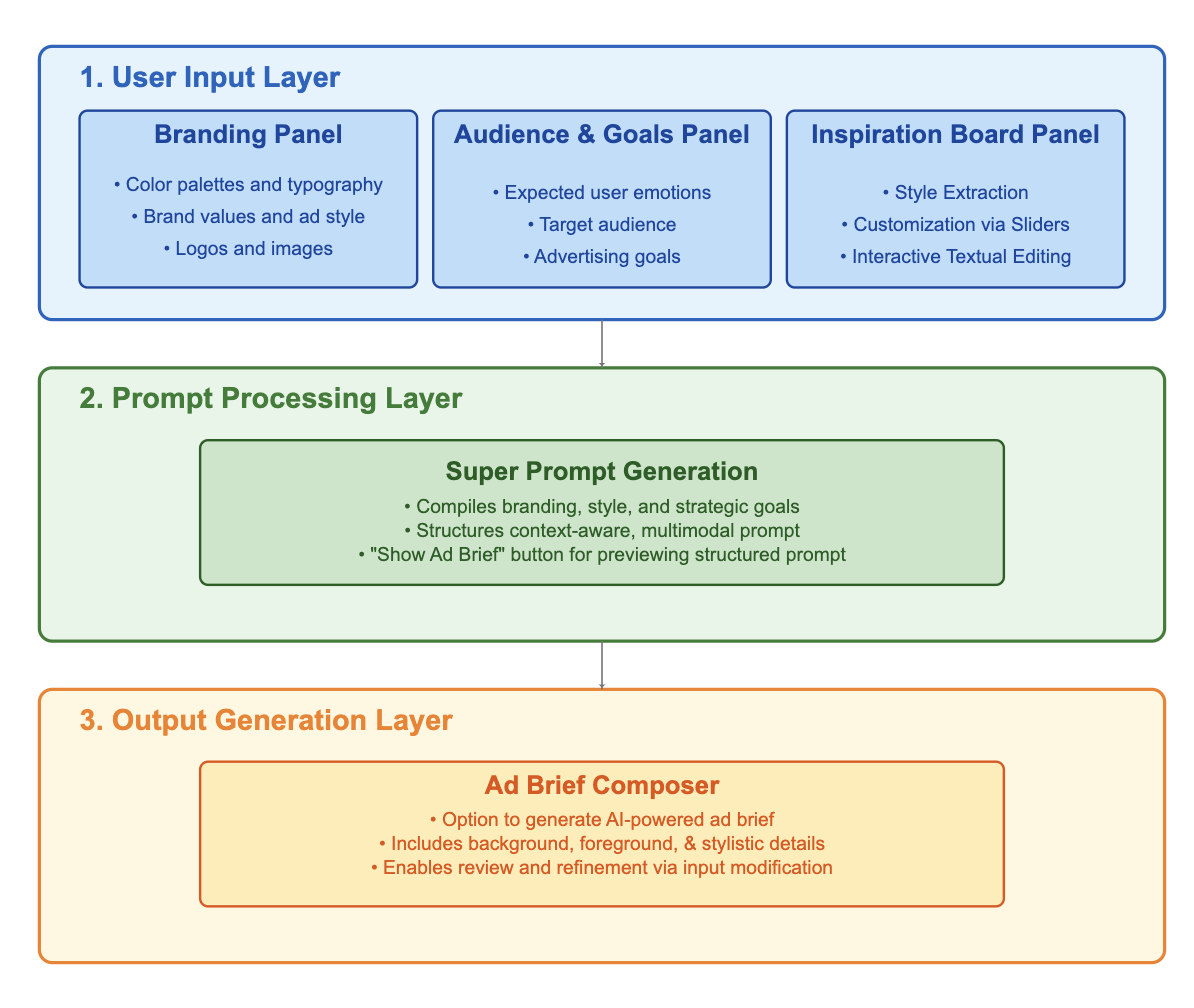} 
    \vspace{-15pt}
    \caption{ACAI Architecture \vspace{-15pt}}
    \label{fig:acai_architecture}
\end{figure}

\subsection{System Overview}
\label{sec:overview}
ACAI leverages Gemini 1.5 Pro, a multimodal large language model (MLLM), to support its generative capabilities. ACAI’s architecture is organized into three core components: the User Input Layer, the Prompt Processing Layer, and the Output Generation Layer (see Figure~\ref{fig:acai_architecture}).

\begin{itemize}
\item[1.] \noindent \textbf{User Input Layer:}
The user input allows users to define brand attributes, advertising objectives, and stylistic preferences through three panels:

\begin{itemize}
    \item \textbf{Branding Panel:} Enables users to specify color palettes, typography, and brand values, as well as upload brand assets (logos, images, key visuals).
    \item \textbf{Audience \& Goals Panel:} Captures advertising objectives, target audience segmentation, and emotional tone preferences to ensure AI-generated Ad brief aligns with marketing strategies.
    \item \textbf{Inspiration Board Panel:} Allows users to upload reference images and interactively extract stylistic elements (e.g., fonts, elements, themes). A Multimodal Large Language Model (MLLM), such as Gemini, analyzes the images to extract regions of interest and generate structured textual descriptors for inclusion in the prompt.
\end{itemize}

By organizing the prompt workflow into guided steps, ACAI supports novice users in articulating their intent more effectively, addressing common challenges in prompt formulation. \cite{subramonyam2024bridging, johnnycantprompt_nonAIexperts}.

\item[2.]\noindent \textbf{Processing Layer:}
In the processing layer, inputs from the three panels are synthesized into a ``super prompt", a structured, multimodal representation of the user’s design intent. This prompt includes both textual attributes (e.g., brand values, objectives) and image-derived descriptors (from the inspiration board). The resulting prompt is processed by the MLLM, which generates a textual Ad Brief designed to reflect the user’s brand identity and marketing goals. 

\item[3.]\noindent \textbf{Output Generation Layer}
This layer incorporates the ``Ad Brief Generator", which converts AI-generated content into a structured brief. The generated Ad Brief contains the following three segments:

\begin{itemize}
    \item \textbf{Summary}: A high-level overview of the advertisement concept.
    \item \textbf{Background}: Visual and contextual details.
    \item \textbf{Foreground}: Key messaging and focal elements.
\end{itemize}

\end{itemize}

The Ad Brief functions as a guide for downstream visual production, whether by a designer or a future generative model. While ACAI does not currently produce visual assets directly, it facilitates human-AI co-creation by maintaining user control over content generation. Users may revise their input fields and regenerate the brief, supporting iterative cycles of ideation and refinement. Future iterations of the system will incorporate input weighting mechanisms to modulate the influence of specific parameters on the generated output. 

\vspace{-5pt}

\begin{figure*}[h]
    \centering
    \includegraphics[width=\linewidth]{
    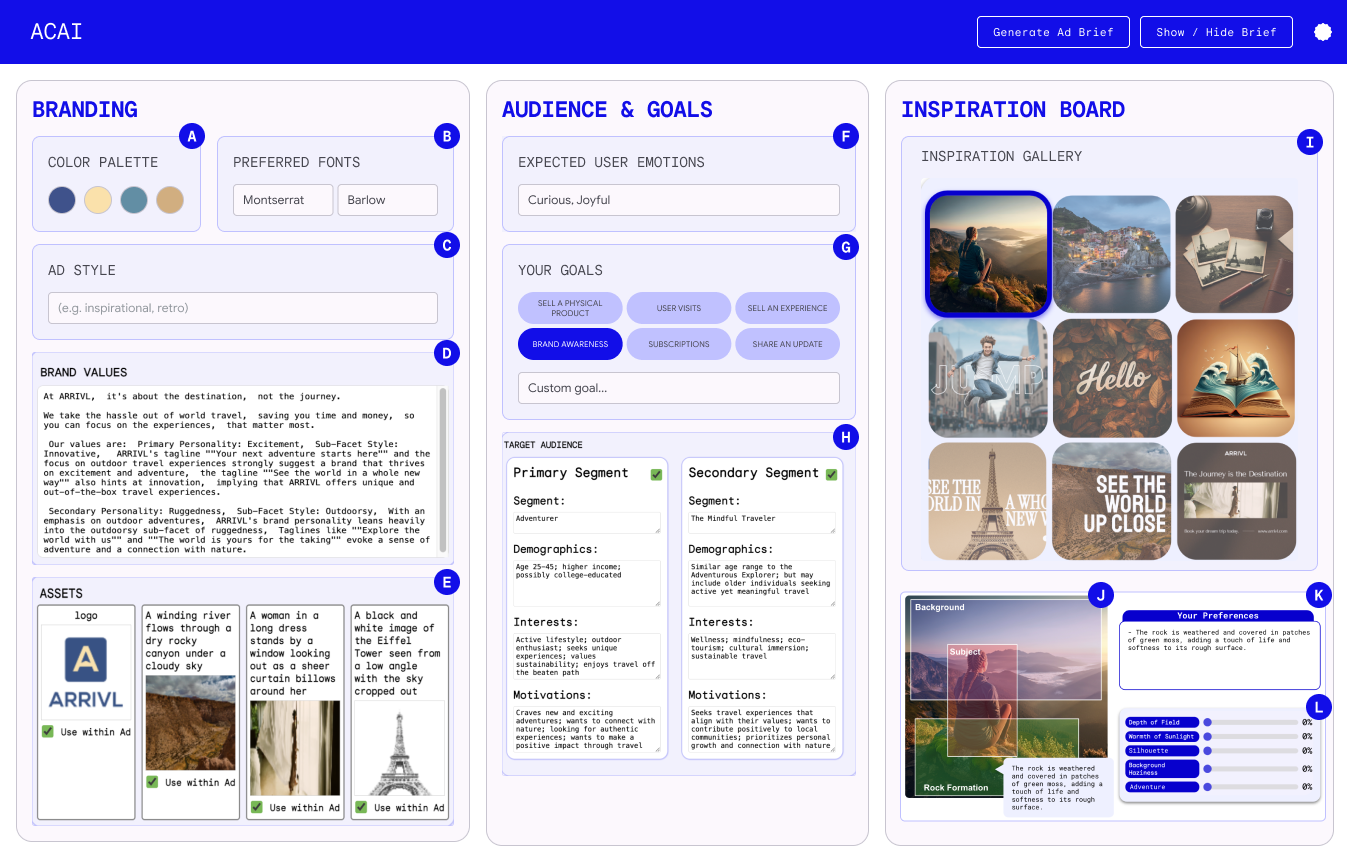} 
    \vspace{-15pt}
    \caption{ACAI Interface featuring the three main panels: Branding (A-E), Audience \& Goals (F-H), and Inspiration Board (I-L) that structure user input for generating brand-aligned advertisement briefs.\vspace{-8pt}}
    \label{fig:acai_interface}
\end{figure*}

\subsection{Implementation walkthrough}
\label{sec:implementation}
To demonstrate how ACAI supports co-creation in real-world settings, we developed a demo brand named Arrivl, a travel agency managed by a hypothetical small business owner, Emily, based in London. Arrivl specializes in providing stress-free travel experiences, helping customers focus on their destination rather than the logistics of planning. To simulate authentic usage, we used a comprehensive set of branding assets for Arrivl, including a defined color palette, brand typography, brand values, logo, and three previously published advertisements. We also specified advertising objectives, target audience segmentation, and intended emotional tone to guide the AI-generated outputs. Additionally, we curated nine reference images for ACAI to analyze, enabling the system to extract relevant stylistic cues (Figure~\ref{fig:acai_interface}). The following section illustrates how ACAI assisted Emily in generating an AI-powered Ad brief.

Emily begins by interacting with the Branding Panel, where she inputs key brand attributes that will shape the advertisement. She selects her brand's colors in the color palette (Figure ~\ref{fig:acai_interface}A), specifies Montserrat and Barlow as the preferred typefaces (Figure ~\ref{fig:acai_interface}B), and uncertain about a preferred visual aesthetic (e.g., retro, natural, or minimalist), Emily chooses not to define a fixed Ad Style (Figure~\ref{fig:acai_interface}C). As a result, ACAI does not apply any stylistic constraints during initial advertisement generation; however, the system remains flexible, allowing Emily to retroactively introduce stylistic preferences at any stage in the co-creation process. This design choice supports customization, even for users who may not have a fully articulated creative vision at the outset. Next, Emily proceeds to the Brand Values (Figure ~\ref{fig:acai_interface}D). While she has the option to input a brand description manually, ACAI automatically generates an initial summary after she uploads brand assets (Figure ~\ref{fig:acai_interface}E) and selects a target audience segment (Figure ~\ref{fig:acai_interface}H). This autogenerated text synthesizes cues from her selected visuals and audience preferences, offering a system-generated interpretation of brand value. The text remains editable both before and after advertisement generation, enabling iterative refinement. This mechanism supports co-creation by lowering the initial barrier of articulating a brand narrative, particularly for novice users or those with limited experience expressing brand values in textual form. Emily then uploads and selects brand assets (Figure ~\ref{fig:acai_interface}E), including Arrivl’s logo and three images from previous advertisements, although other types of visual material may also be used. While ACAI allows users to select between zero and four assets, Emily opts to include all four. These selected assets are treated as priority inputs and prominently featured in the generated Ad brief, ensuring that user-provided content substantively informs the system’s output.

With Arrivl's core branding traits set, Emily moves onto Audience \& Goals Panel. Given Arrivl's focus on travel, Emily enters `curious' and `joyful' as the expected user emotions (Figure~\ref{fig:acai_interface}F). Next, a range of advertisement goals (Figure~\ref{fig:acai_interface}G) are available for selection as well as a text field for a custom goal. Emily selects `brand awareness' as the goal for both Arrivl's primary and secondary audiences (Figure~\ref{fig:acai_interface}H). To explore potential market segments, Emily interacts with the Target Audience panel (Figure~\ref{fig:acai_interface}H), entering “Adventurer” and “The Mindful Traveler” as her primary and secondary segments. Drawing on these entries, along with information from other ACAI panels that is either manually input by Emily or automatically generated, ACAI produces corresponding demographic traits, interests, and motivations. This feature is designed to help novice users articulate their audience strategy and offers opportunities for further refinement.

In the final stage, Emily interacts with the Inspiration Board (Figure~\ref{fig:acai_interface}I), where she engages with visual references to articulate her desired aesthetic. The interface displays a gallery of nine images (Figure~\ref{fig:acai_interface}I), which may include prior advertisements or aspirational references. Emily selects the top-left image from the gallery. Once selected, the image appears in a dedicated workspace beneath the gallery. ACAI’s underlying model, Gemini, analyzes the selected image and segments it into contextually salient visual components, such as the subject, background, and other prominent elements (in this case, a rock formation; (Figure~\ref{fig:acai_interface}J). Hovering over any segmented region reveals design-oriented annotations, which are natural language descriptions that articulate the stylistic or compositional qualities of the selected element. When Emily hovers over the rock formation, for example, the system surfaces the annotation: “weathered and covered in patches of moss, giving a sense of life and softness to its rough surface” (Figure~\ref{fig:acai_interface}K). These annotations assist novice designers like Emily in interpreting the aesthetic tone of the image and help inform decisions about which visual attributes to incorporate into the advertisement. To support finer-grained control, ACAI dynamically generates a set of adjustable sliders corresponding to image-specific features such as Depth of Field, Source of Light, Silhouette, Backdrop, and Adventure (Figure~\ref{fig:acai_interface}L).

\section{Discussion}
This paper contributes to ongoing HCI discourse on the promptability challenges faced by novice users in generative AI systems~\cite{subramonyam2024bridging,johnnycantprompt_nonAIexperts}. Grounded in formative research with small business owners (SBOs) in the United Kingdom, we designed ACAI to support them in creating brand-aligned advertisements through structured prompting and multimodal interface interaction. Below, we reflect on ACAI's broader implications for reimagining generative workflows and moving toward inclusive AI interface design.

\subsection{Reimagining generative workflows}
ACAI reimagines how novice users engage with generative AI by moving beyond conventional free-form text prompting toward a structured, multimodal interaction paradigm. Unlike systems that rely primarily on open-ended textual input, ACAI enables users to steer content generation through clearly defined input panels and diverse input modalities. It is designed to help small business owners create brand-aligned content through an interface that reduces the cognitive effort required for prompt formulation.

By offering a more expressive and guided alternative to traditional prompting, ACAI highlights broader design implications related to interaction flexibility, editability, and promptability. These are critical considerations in the development of evolving AI-augmented UI workflows. While ACAI does not generate UI layouts directly, it contributes to the workshop’s central theme by demonstrating how AI functionality can inform interface architecture, input design, and creative tooling practices.

ACAI’s interface, described in detail in Section~\ref{sec:implementation}, illustrates how structured input panels, multimodal selection mechanisms, visual style extraction, and dynamically generated sliders enable users to adjust visual attributes such as mood, spatial depth, and composition. These affordances reflect a broader reimagining of how generative systems can support control, co-creative flexibility, and intent alignment within creative workflows.

.

\subsection{Toward Inclusive AI Interface Design}
Through this work, we offer a perspective on how designing AI-powered interfaces for novice creators can inform the development of more inclusive interaction paradigms. We highlight how ACAI’s structured and multimodal interface provides alternatives to dominant text-based prompting approaches, thereby expanding the design space for generative AI user interface development. In articulating these possibilities, we aim to provoke broader reflection on how generative AI tooling might move beyond expert-centric workflows toward more scaffolded and equitable interaction paradigms.

\section{Conclusion and Future Work}
In this paper, we introduced ACAI (AI Co-Creation for Advertising and Inspiration), a multimodal generative AI prototype designed to support novice designers through a structured alternative to traditional prompt-based interfaces. Our formative study with small business owners revealed three persistent limitations in current generative workflows: difficulty in externalizing brand intuition through text, limited opportunities for fine-grained control during and after content generation, and the frequent production of generic outputs misaligned with brand identity. ACAI addresses these challenges through a panel-based interface that enables users to specify branding elements, define audience goals, and select visual inspirations, facilitating more directed and brand-consistent co-creation. Our approach demonstrates how structured interfaces can foreground user-defined context to improve both alignment and promptability in novice workflows. ACAI contributes to ongoing HCI research on co-creative tooling by offering an accessible, context-aware alternative to conventional prompt-based interaction ~paradigms. We refer readers to~\cite{karnatak2025acaisbosaicocreation} for a full discussion of the evaluation, which investigates how ACAI supports small business owners in creating brand-aligned advertisements within situated design workflows. Looking ahead, future research could explore how ACAI’s co-creation model might be adapted to other settings that require contextual grounding but involve users with limited design expertise.

\bibliographystyle{ACM-Reference-Format}
\bibliography{bibliography.bib}

\end{document}